\documentclass{aa}
\usepackage{graphics}
\usepackage{rotating}   
\usepackage{psfig}
\begin{document}

   \thesaurus{01     % A&A Section 03
              (11.01.2;  % Galaxies: active,
               11.09.1 III Zw 2; % Galaxies: individual
               11.10.1;  % Galaxies: jets,
               11.19.1)}  % Galaxies: Seyfert,

   \title{III~Zw~2, the first superluminal jet in a Seyfert galaxy}

%   \subtitle{}

   \author{A. Brunthaler \inst{1}
   \and    H. Falcke \inst{1}
   \and    G. C. Bower \inst{2}
   \and    M. F. Aller \inst{3}
   \and    H. D. Aller \inst{3}
   \and    H. Ter\"asranta \inst{4}
   \and    A.P. Lobanov \inst{1}
   \and    T.P. Krichbaum \inst{1}
   \and    A.R. Patnaik \inst{1}
          }

   \offprints{hfalcke@mpifr-bonn.mpg.de}

   \institute{Max-Planck-Institut f\"ur Radioastronomie,
              Auf dem H\"ugel 69, 53121 Bonn, Germany
   \and       National Radio Astronomy Observatory, 
              P.O. Box O, Socorro, NM 87801-0387, USA
   \and       Astronomy Department, University of Michigan,
              Ann Arbor, MI 48109-1090, USA
   \and       Mets\"ahovi Radio Observatory, 
              Metsahovintie, FIN-02540 Kylm\"al\"a, Finland
             }

   \date{Astronomy \& Astrophysics Letters, in press}

   \maketitle

   \begin{abstract} So far all relativistically boosted jets with
   superluminal motion have only been detected in typical radio
   galaxies with early type host galaxies. We have now discovered
   superluminal motion in the Seyfert I galaxy III~Zw~2, classified as
   a spiral. Superluminal motion was first inferred from the spectral
   evolution of the source and then confirmed by VLBI
   observations. The lower limit for the apparent expansion speed is
   $1.25\pm0.09$~c. The fact that the spectral and spatial evolution
   are closely linked demonstrates that we are dealing with real
   physical expansion. Prior to this rapid expansion we have seen a
   period of virtually no expansion with an expansion speed less than
   0.04 c. Since III~Zw~2 is also part of a sample of so called
   radio-intermediate quasars (RIQ), it confirms earlier predictions
   of superluminal motion for this source, based on the argument that
   RIQs could be relativistically boosted jets in radio-weak quasars
   and Seyfert galaxies.

   \end{abstract}

%
%________________________________________________________________

\section{Introduction}

The radio properties of quasars with otherwise very similar optical
properties can be markedly different. There is a clear dichotomy
between radio-loud and radio-quiet quasars in optically selected
samples. The radio-loudness is usually characterized by the
radio-to-optical flux ratio. In the PG quasar sample, which is
probably the best studied quasar sample in the radio and optical
(Kellermann et al.~\cite{kellermann}, Boroson
\& Green \cite{boroson}), radio-loud and radio-weak quasars separate
cleanly in two distinct populations (e.g.~Kellermann et al.~1989).

It is known that radio-loud AGN almost never reside in late type,
i.e.~spiral galaxies (e.g.~Kirhakos et al.~\cite{kirhakos}, Bahcall et
al.~\cite{bahcall}) whereas radio-quiet quasars appear both in spiral
and in elliptical host galaxies.  Furthermore, all relativistically
boosted jets with superluminal motion and typical blazars have been
detected in early type galaxies (e.g. Scarpa et al.~\cite{scarpa}).
It is still unclear, why AGN in spiral galaxies, at the same optical
luminosity as their elliptical counterparts, should not be able to
produce the powerful, relativistic jets seen in radio galaxies.

However, a few sources with intermediate radio-to-optical ratios
appear to be neither radio-loud nor radio-quiet.  They form a distinct
subclass with very similar radio morphological and spectral
properties.  They all have a compact core at VLA scales and a flat and
variable spectrum in common. These properties are very similar to the
ones of radio cores in radio-loud quasars, but their low
radio-to-optical ratio and their low extended steep-spectrum emission
is atypical for radio-loud quasars.  Miller et al.~(\cite{miller}) and
Falcke et al.~(\cite{falcke95},
\cite{falcke96a}\&b) have identified a number of these sources, called
``radio-intermediate quasars'' (RIQs), and suggested that they might
be relativistically boosted radio-weak quasars or ``radio-weak
blazars''. This would imply that most, if not all, radio-quiet quasars
also have relativistic jets. In fact, VLBI observations of radio-quiet
quasars already have shown high-brightness temperature radio cores and
jets (Blundell \& Beasly \cite{blundell}). A crucial test of the
relativistic jet hypothesis is the search for apparent superluminal
motion in these sources. A prime candidate for detecting this is the
brightest radio source in the RIQ sample, III~Zw~2, which we discuss
in this paper.

III~Zw~2 (PG 0007+106, Mrk 1501, $z=0.089$) is one of the most extremely 
variable radio sources and a very unusual AGN. It was discovered by Zwicky 
(\cite{zwicky}), classified as a Seyfert I galaxy (e.g., Arp 
\cite{arp}; Khachikian \& Weedman \cite{khachikian}; Osterbrock 
\cite{osterbrock}), and later also included in the PG quasar sample 
(Schmidt \& Green \cite{schmidt}). The host galaxy was classified as a
spiral (e.g. Hutchings \& Campbell \cite{hutchings_campbell}) and a
spiral arm was claimed (Hutchings \cite{hutchings}). A disk model was
later confirmed by fitting of model isophotes to near-IR images
(Taylor et al.~\cite{taylor}).

The most interesting property of III~Zw~2, however,  is its extreme
variability at radio and other wavelengths with at least 20-fold
increases in radio flux density within 4 years (Aller et
al.~\cite{aller}). The source also shows optical (Lloyd \cite{lloyd})
and X-ray variability (Kaastra \& de Korte \cite{kaastra}; Pounds
\cite{pounds}).

III~Zw~2 is a core-dominated flat-spectrum AGN with only a faint
extended structure (see Unger et al.~\cite{unger}). The weak extended
radio emission and the host galaxy is quite typical for a Seyfert
galaxy. Its [O{\sc III}] luminosity is a mere factor three brighter
than that of a bright Seyfert galaxy like Mrk~3 (e.g.~Alonso-Herrero
et al.~\cite{alonso}) which explains why it has been classified as
either a Seyfert galaxy or a quasar. In this luminosity region a
distinction between the two may not be of much significance.

Earlier VLBI observations of the source have only shown a
high-brightness temperature core (Falcke et al.~\cite{falcke96b},
Kellermann et al.~\cite{kellermann98}) and recent Millimeter-VLBI
observations by Falcke et al.~(\cite{falcke99}) just barely resolved
the source into two very compact components.

Based on its average optical-to-radio ratio of $\sim200$ and its radio
properties, it was suggested that III~Zw~2 could be a
radio-intermediate quasar and the presence of superluminal motion in
this source was predicted. Here we will discuss multi-epoch VLA and
millimeter-VLBI observations of this source confirming this
hypothesis.

%__________________________________________________________________

\section{Observations and Data Reduction}

%                                     Two column figure (place early!)
%______________________________________________ Gamma_1 (lg rho, lg e)

%

In 1997 we detected the onset of a new major radio outburst in
III~Zw~2 and we initiated a target of opportunity program to monitor
the spectral evolution of the burst with the VLA and its structural
evolution with the VLBA.

The VLA observations were made at six frequencies ranging from 1.4 GHz
to 43 GHz in A, B, C and D configuration from 1998 September until
now. The three epochs discussed in this paper where obtained on 1998
November 04, 1999 March 23 and 1999 July 07 in CnB, D and A
configuration respectively. The source 3C48 was used as the primary
flux density calibrator, and III~Zw~2 was self-calibrated and mapped
with the Astronomical Image Processing System (AIPS).  Since the 1.4
GHz observation in March 1999 was heavily confused by the nearby sun,
we estimated the flux density by interpolation between earlier and
later epochs.

We observed III~Zw~2 with the VLBA on 1998 February 16, 1998 June 13,
1998 September 22, 1998 December 12, and 1999 July 15 at 43 and 15
GHz. We used a total bandwidth of 64 MHz for a full 8 hour scan,
except for the last observation which was a 4 hour scan. We spent
three-quarters of the available observing time at 43 GHz and one
quarter of it at 15 GHz.  For the second epoch, we used the Effelsberg
100 m telescope in combination with the VLBA. We reduced the data
using the software packages AIPS and DIFMAP (Shepherd, Pearson, \&
Taylor~\cite{shepherd}). Fringes were detected in the III~Zw~2 data on
all baselines.  We calibrated the gains using system temperature
information and applied atmospheric opacity corrections. To obtain a
reliable total flux estimate, amplitude gains for stations with bad
weather conditions were scaled up to match the other antennas. The
data were then self-calibrated, first using phase-only and later
phase-amplitude self-calibration with solution intervals slowly
decreasing down to one minute. A more detailed discussion of the VLA
spectra and the 15 GHz VLBA-data will be presented in a forthcoming
paper.

\begin{figure}
 \resizebox{\hsize}{!}{\includegraphics[bbllx=2.2cm,bburx=19.6cm,bblly=2cm,bbury=26.7cm,clip=,angle=-90]{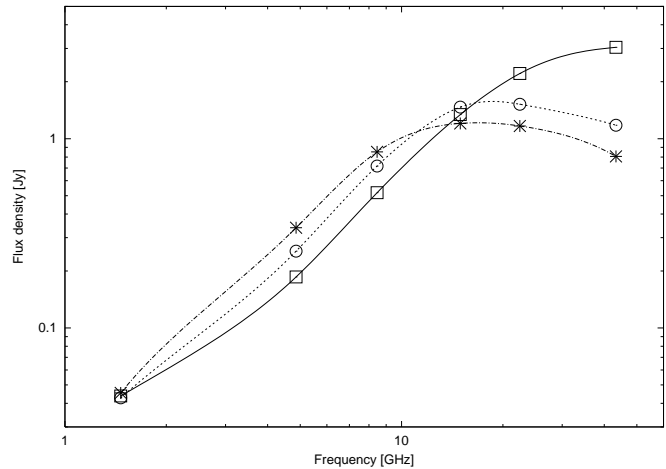}}
 \caption{VLA spectra of III~Zw~2 from 1998 November (boxes), 1999 March (circles) 
 and 1999 July (stars). The peak frequency dropped quickly within a few months 
 from 43 GHz to 15 GHz. The lines are concatenations of the points and show the 
 smoothness of the spectra.}
 \label{spec}
\end{figure}

%______________________________________________________________

\section{Results}

The initial spectrum of III~Zw~2 from 1998 May was presented in Falcke
et al.~(\cite{falcke99}). It was highly inverted at centimeter
wavelengths with a spectral index of $\alpha = +1.9\pm0.1$ between 4.8
and 10.5 GHz. The entire outburst spectrum from 1.4 GHz to 666 GHz
could basically be fitted by only two homogeneous, synchrotron
components which are optically thin at high frequencies and become
self-absorbed below 43 GHz. This spectral turnover frequency stayed
constant until 1998 November and hence we expected no strong
structural change during this time (Fig.~\ref{spec}).

\begin{figure}
 \resizebox{\hsize}{!}{\includegraphics[bbllx=4.6cm,bburx=19.3cm,bblly=9.1cm,bbury=23.4cm,clip=]{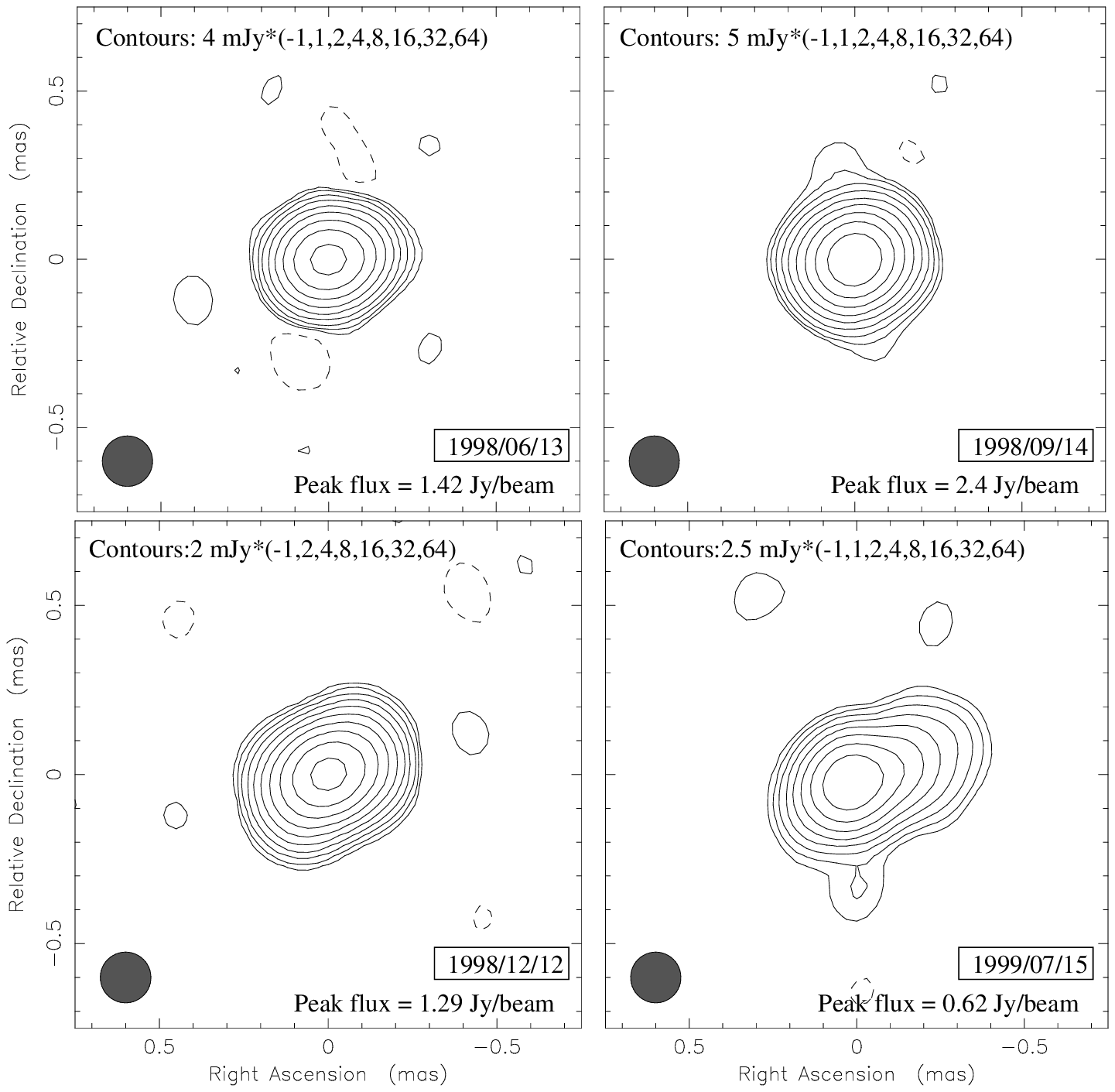}}
 \caption{Four epochs of VLBA maps of III~Zw~2 at 43 GHz convolved with a superresolved beam of 150 $\mu$as. The original beam sizes were $0.29\times 0.12$ mas at a position angle (P.A.) of $-5^{\circ}$ in June 1998, $0.31\times 0.16$ at a P.A. of $11^{\circ}$ in September 1998, $0.38\times0.17$ at a P.A. of $-4^{\circ}$ in December 1998 and $0.5\times0.14$ at a P.A. of $-18^{\circ}$ in July 1999.}
\label{vlba2}
\end{figure}

Our first three VLBA epochs were made while the spectral peak was
constant at 43~GHz. The core itself is resolved at 43 GHz at all
epochs. To represent the extent of the source, the non-zero closure
phases at long baselines for the 43 GHz data were fit by two
point-like components.  A rough estimate of the formal statistical
errors of the component separation was obtained by dividing the
original beam size by the post-modelfit signal-to-noise ratio
(e.g. Fomalont~\cite{fomalont}, Sect. 2.3). The errors were of the
order of $1\mu as$ for the first four epochs and $10\mu as$ for the
fifth epoch. Additional to this very small statistical error, there
should be a larger systematic error which is difficult to quantify.
To minimize this systematic error we used very similar reduction
procedures for each epoch.

In accordance to the VLA data, the maps of the first three epochs show
no structural change and the separation of the fitted components
stayed constant within the statistical errors at $\sim$76$ \pm
2~\mu$as corresponding to $\sim$0.11~pc (see Table~\ref{table}) for an
angular size distance of 307.4 Mpc ($H_{0}=75$ km/sec/Mpc, $q_{0}=0.5$
as used in this paper).

After 1998 November, the VLA observations showed a dramatic change in
the spectrum.  The spectral peak dropped quickly to 15~GHz within a
few months (Fig.~\ref{spec}). Since the peak in this source is caused
by synchrotron self-absorption (Falcke et al.~\cite{falcke99}), the
fast change in peak frequency implied a similarly strong morphological
change, i.e. a rapid expansion. To roughly estimate the expected
expansion speed, we applied a simple equipartition jet model with a $R
\propto \nu_{\mathrm{ssa}}^{-1}$ dependence (e.g. Blandford \& K\"onigl
\cite{blandford_koenigl}; Falcke \& Biermann \cite{falcke_biermann}). For an 
initial source size $R = 0.11$~pc and a self-absorption frequency
$\nu_{\mathrm{ssa}}=43$~GHz in 1998 November, we calculate a source size of
0.32~pc for a self-absorption frequency of 15~GHz in 1999 March. Thus
we predicted an apparent expansion speed of 1.9~c after the correction
for cosmological time dilatation and asked for further
VLBA-observations.

\begin{figure}
 \resizebox{\hsize}{!}{\includegraphics[angle=-90]{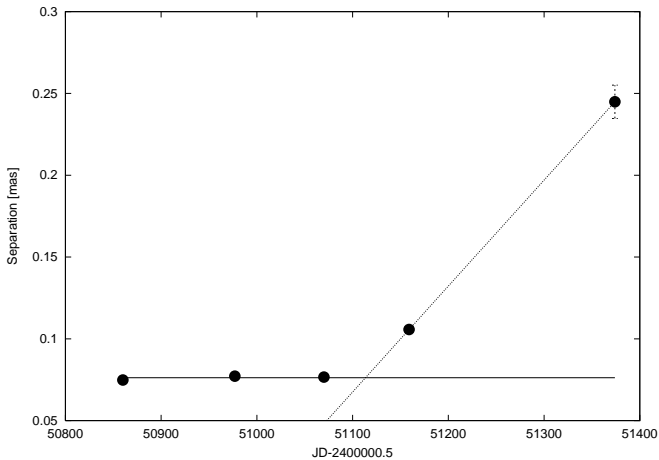}}
 \caption{Component separation from model fitting of point-like
 components to the closure phases and amplitudes at 43 GHz. The statistic 
 errors for the first four epoch are smaller than the symbols and should be
 dominated by systematic errors. The
 separation of the first three epochs is consistent with an expansion speed 
 $\le0.04~c$ (solid line). The expansion speed between the
 fourth and fifth epoch is $1.25\pm0.09~c$.}  \label{move7}
\end{figure}

Indeed the fifth epoch of VLBA observations showed a dramatic
structural change compared to the earlier epochs (Fig.~\ref{vlba2}). A
model of at least three point-like components is required now. It was
not possible to fit the closure phases with a two-component model as
in the earlier epochs or to get rid of this third component during
self-calibration. The separation of the outer components for all five
epochs is plotted in Fig.~\ref{move7}. While the separation at the
first three epochs is consistent with an expansion speed of $\le
0.04~c$, the fifth epoch shows a rapid expansion.  The apparent
expansion speed between the outer components in the 4th and 5th epoch
is $1.25\pm0.09~c$.

This value is only a lower limit and increases to 2.66~c if one
considers the time range from December until March during which most
of the spectral evolution occurred.  Applying the standard equation
for superluminal motion, $\beta_{\mathrm{app}}=\frac{\beta \sin \theta}{1-\beta
\cos\theta}$ (e.g. Krolik \cite{krolik}), to a value of $\beta_{\mathrm{app}}=2.66$ 
constrains the 
maximal angle between the jet and the line of sight to
$\theta=41^{\circ}$, since $\beta<1$.

   \begin{table}
      \caption[]{Separation D and position angle P.A. of the 
        outermost point-like components of our model-fits to the uv-data.}
         \label{table}
      \[
  \begin{tabular}{p{0.18\linewidth}|cp{0.14\linewidth}cp{0.10\linewidth}cp{0.10\linewidth}cp{0.10\linewidth}c}
           \hline
 Date & Total flux [Jy] &  D [mas] & D [pc] & P.A.\\
            \hline
	   1998/02/16 & 1.54 $\pm$ 0.15 & 0.075 & 0.11 & $-84^{\circ}$\\
           1998/06/13 & 1.69 $\pm$ 0.17 & 0.077 & 0.11 & $-78^{\circ}$\\
           1998/09/14 & 2.92 $\pm$ 0.29 & 0.077 & 0.11 & $-72^{\circ}$\\
           1998/12/12 & 1.76 $\pm$ 0.18 & 0.106 & 0.16 & $-63^{\circ}$\\
           1999/07/15 & 0.92 $\pm$ 0.09 & 0.245 & 0.37 & $-71^{\circ}$\\
            \hline
         \end{tabular}
      \]
   \end{table}

%______________________________________________________________

\section{Summary \& Discussion}

We found a close connection between the spectral and structural
evolution of the radio outburst in III~Zw~2.  While the
self-absorption frequency remained constant, we observed no change on
VLBA-scales. The excellent consistency of the relative sizes during
the first three epochs on the few micro-arcsecond level demonstrates
the good quality of our data. The quick drop in peak frequency after
1998 November marked the beginning of a strong structural change. The
spectral peak dropped from 43 GHz to 15 GHz within a few months and
the VLBA-maps show a rapid expansion with an apparent expansion
velocity of 1.25 c. The fact that spectral and spatial evolution 
are so closely linked also demonstrates that we are dealing
with real physical expansion.

With its faint extended radio structure and its spiral host galaxy
III~Zw~2 is clearly not a radio-loud quasar, but has properties very
typical of luminous Seyfert galaxies or radio-quiet quasars.  The
detection of superluminal motion in this galaxy now clearly shows that
we are dealing with a relativistic jet on sub-pc scales.  To our
knowledge this is the first detection of superluminal motion in a
spiral galaxy with a Seyfert nucleus. The maximum aspect angle of
$41^{\circ}$ from the superluminal motion (see Sec.~3) is in good
agreement with the orientation based unified scheme (e.g. Antonucci
\cite{antonucci}) for AGN where Seyfert I galaxies are seen under
intermediate or small aspect angles, so that the nucleus is not
obscured by a dusty torus.

For the question of the nature of the radio-loud/radio-quiet dichotomy
this means that radio-weak and radio-loud quasars can indeed have
central engines that are in many respects very similar. Their optical
properties are almost indistinguishable and both types of quasars can
produce relativistic jets in their nuclei. The finding of superluminal
motion supports the hypothesis of Miller et al.~(1993) and Falcke et
al.~(\cite{falcke96a}) that RIQs are relativistically boosted
intrinsically radio-weak AGN.

While these general conclusions seem to be fairly robust, some
characteristics of the outburst need to be modeled in more
detail. The initial phase of the flux density rise with its
millimeter-peaked spectrum and no detectable expansion perhaps has to
be explained similar to the physics of Gigahertz-Peaked-Spectrum (GPS)
sources, i.e.~ultra-compact hotspots pumped up and powered by a jet
interacting with the interstellar medium or the torus. The rapid
expansion thereafter could have marked the phase where the jet breaks
free and starts to propagate relativistically into a lower-density
medium.  Another explanation of the initial phase could be a fast jet
moving through quasi-stationary components as proposed for sources
like 4C\,39.25 (e.g. Alberdi et al.~\cite{alberdi}). In any case, this
dramatic structural change should go together with a change of the
polarization vector. This must be checked by future experiments.

The initial slow expansion has also possible implications for the
interpretation of other Seyfert galaxies. For example, in observations
of the Seyfert galaxies Mrk 348 and Mrk 231 (Ulvestad et
al.~\cite{ulvestad}) only sub-relativistic expansion was found. Hence,
one could raise the question whether the region of relativistic
expansion in these two sources is at even smaller scales or whether
they just happened to be in a `slow' phase, similar to III~Zw~2
early on. If true, it is still possible that Seyfert jets are launched
relativistically but are slowed down and disrupted significantly
already on the sub-parsec scale. Therefore it would be important to
follow the spectral evolution of these sources in more detail.

\begin{acknowledgements}
We acknowledge the help of an anonymous referee. This research was
supported by DFG grants Fa~358/1-1 and 358/1-2. The National Radio
Astronomy Observatory is a facility of the National Science Foundation
operated under cooperative agreement by Associated Universities, Inc.
\end{acknowledgements}

\clearpage

\end{document}